\documentstyle[12pt]{article}
\def\beq{\begin{equation}}
\def\eeq{\end{equation}}
\def\bea{\begin{eqnarray}}
\def\eea{\end{eqnarray}}
\def\bq{\begin{quote}}
\def\eq{\end{quote}}

\def\NP{{\it Nucl.Phys.} }
\def\PL{{\it Phys.Lett.} }
\def\PR{{\it Phys.Rev.} }

\def\ZP{{\it Z.Phys.} }

\def\gappeq{\mathrel{\rlap {\raise.5ex\hbox{$>$}}
{\lower.5ex\hbox{$\sim$}}}}

\def\lappeq{\mathrel{\rlap{\raise.5ex\hbox{$<$}}
{\lower.5ex\hbox{$\sim$}}}}

\begin{document}
\pagestyle{empty}
\begin{flushright}
{CERN-TH/96-161}
\end{flushright}
\vspace*{5mm}
\begin{center}
{\bf NUCLEON SPIN STRUCTURE,  TOPOLOGICAL SUSCEPTIBILITY} \\
{\bf AND THE $\eta^\prime$ SINGLET AXIAL VECTOR COUPLING }$^{*)}$ \\
\vspace*{1cm} 
{\bf G. Grunberg} \\
\vspace{0.3cm}
Theoretical Physics Division, CERN \\
CH - 1211 Geneva 23 \\
and \\
Centre de Physique Th\'eorique de l'Ecole Polytechnique$^{+)}$\\
F - 91128 Palaiseau Cedex \\
{\tt e-mail: grunberg@orphee.polytechnique.fr}
\vspace*{2cm}\\  
{\bf ABSTRACT} \\ \end{center}
\vspace*{5mm}
\noindent
The observed small value of the first moment of the polarized nucleon 
spin structure function $g_1$ may be interpreted, in the Veneziano--Shore
approach, as
a suppression of the first moment $\chi^\prime(0)$ of the QCD topological
susceptibility. I give an extension of the Witten--Veneziano argument for
the $U(1)$
problem, which yields the $O(1/N)$ correction to the $N = \infty$ relation
$\chi^\prime(0)/F^2_0 = 1$ (where $F_0$ is the $\eta^\prime$ axial vector
coupling).
The correction, although negative, seems too small to account for the data. I
further argue that the $(\eta , \eta^\prime)\rightarrow\gamma\gamma$ and
$J/\psi\rightarrow (\eta , \eta^\prime)\gamma$ decays indicate an enhancement
rather than a suppression of $F_0$. A substantial gluon-like contribution in
$\langle 0\vert\partial^\mu j^{(0)}_{\mu_5} \vert\gamma\gamma \rangle
\vert_{q^2=0}$,
which could parallel a similar one  in the corresponding nucleon matrix element,
is suggested.

\vspace*{1cm} 
\noindent 
\rule[.1in]{16.5cm}{.002in}

\noindent
$^{*)}$ Researched supported in part by the EC Program ``Human Capital and
Mobility", Network ``Physics at High-Energy Colliders", contract CHRX-CT93-0357.

\noindent
$^{+)}$ C.N.R.S. UPRA 0014 
\vspace*{0.5cm}

\begin{flushleft} CERN-TH/96-161 \\
June 1996
\end{flushleft}
\vfill\eject

\setcounter{page}{1}
\pagestyle{plain}

\section{Introduction}

The experimental discovery of a substantial suppression of the first moment
$\Gamma
^{(p,n)}_1$ of the polarized nucleon structure function with respect to the
``na\"\i ve"
(Ellis--Jaffe) OZI limit prediction \cite{aaa} has spurred a lot of theoretical
interest (for reviews, see Refs. \cite{bb}, \cite{cc}) in recent years.
Specifically, consider the full QCD expression \cite{dd} for
$\Gamma_1^{(p,n)}$ (in
the $\overline{\rm MS}$ scheme with $N_f = 3$):
\bea
\Gamma_1^{(p,n)}(Q^2) &=& \int^1_0 dx~g^{(p,n)}_1(x,Q^2) = {1\over 6} \left(\pm
G_A^{(3)} + {1\over\sqrt{3}} G_A^{(8)}\right)  \nonumber \\
&& 
\left(1 - {\alpha_s\over\pi} - 3.58
\left({\alpha_s\over\pi}\right)^2 - 20.22 \left({\alpha_s\over\pi}\right)^3 +
\ldots\right) \nonumber \\
&&+
{1\over 9} G_{A,inv}^{(0)} \left( 1 - {1\over 3}~{\alpha_s\over\pi} - 0.550
\left({\alpha_s\over\pi}\right)^2 + \ldots \right)
\label{1}
\eea
where $\alpha_s = \alpha_s(Q)$ and $G_A^{(i)}$ are zero-momentum transfer form
factors in the proton matrix element of the axial vector currents:
$\langle p\vert j^{(i)}_{\mu_5}\vert p\rangle =   G_A^{(i)} \bar p
\gamma_\mu\gamma_5 p
+ 
\ldots$. I stress that the radiative corrections in Eq. (\ref{1}) imply that the
$\underline{\rm scale~independent}$, renormalization group--invariant
singlet axial
vector current $j^{(0)}_{\mu_5 , inv}$ has been used to define
$G_{A,inv}^{(0)}$. It is obtained from the current $j^{(0)}_{\mu_5}(\mu)$
renormalized in the standard way at scale $\mu$ in the  $\overline{\rm MS}$
scheme
by factorizing out the anomalous dimension factor generated by the $U(1)$
anomaly:
\beq
j^{(0)}_{\mu_5}(\mu) =  j^{(0)}_{\mu_5,inv} ~~\bigg[ 1 + 0
(\alpha_s(\mu))\bigg]~.
\label{2}
\eeq
The crucial feature of Eq. (\ref{2}) is that $j^{(0)}_{\mu_5}(\mu)$ has a
``parton
model" $(\mu\rightarrow\infty, \alpha_s(\mu)\rightarrow 0) $ limit, owing to the
special feature that the anomalous dimension starts only at $O(\alpha_s^2)$. It
follows that $G_{A,inv}^{(0)}$ (also denoted as $\Delta\Sigma_{inv}$ \cite{dd}
or $\Delta\Sigma_{\infty}$ \cite{bb}) is a $\underline{\rm physical}$,
$\underline{\rm \mu -independent}$ constant, which stands on the $\underline{\rm
same~footing}$ as $G^{(3)}_A$ and $G^{(8)}_A$, and that the whole physical
$Q^2$ dependence is entirely contained in the (renormalization group--invariant)
series in $\alpha_s(Q)$ in Eq.~(\ref{1}). (That $G_{A,inv}^{(0)}$ is a physical
parameter should be clear from the observation that, at $Q^2 = \infty$, Eq.
(\ref{1}) gives the parton-model-like sum rule: $\Gamma_1^{(p,n)}(Q^2 =
\infty) =
{1\over 6} (\pm G^{(3)}_A + {1\over\sqrt{3}} G^{(8)}_A) + {1\over 9}
G^{(0)}_{A,inv}$). Experimentally, one finds \cite{cc} $G^{(0)}_{A,inv}
\simeq 0.25$ (where I have taken into account the radiative corrections), to be
compared with the Ellis--Jaffe value $G^{(0)}_{A\vert OZI} \simeq 0.58$, i.e. 
$G^{(0)}_{A,inv}/G^{(0)}_{A\vert OZI} \simeq 0.43$, roughly a factor of 2.

An interesting proposal to understand this suppression has been put forward in
Ref. \cite{ee}, where it has been suggested that it may be a
(target-independent)
effect related to the first moment of the QCD topological susceptibility
$\chi(q^2)$, namely (for three flavours):
\beq
{G^{(0)}_{A,inv}\over G^{(0)}_{A\vert OZI}} \simeq  {\sqrt{\chi^\prime (0)}\over
{F_\pi\over \sqrt{6}}}
\label{3}
\eeq
where $F_\pi$ = 93 MeV,
\beq
\chi(q^2) \equiv \int d^4x~e^{iq.x} \langle 0\vert T^*\bigg(Q_{inv}(x)
Q_{inv}(0)\bigg)\vert 0
\rangle
\label{4}
\eeq
and $Q_{inv}(x)$ is the anomalous divergence of the singlet axial vector
current:
\beq
\partial^\mu j^{(0)}_{\mu_5,inv} = 3 {\alpha_s\over 4\pi} F\tilde F \equiv
6 Q_{inv}~.
\label{5}
\eeq
The basic physical assumption behind Eq. (\ref{3}) is that large Zweig rule
violations in $G^{(0)}_{A,inv}$ are to be found mainly in the
$\sqrt{\chi^\prime(0)}$ factor, which embodies the typical $q\bar q \rightarrow
2$-gluons annihilation diagrams, which are supposed to most strongly violate the
Zweig rule. In this note, I examine new ways to test this assumption. In
the next
section, I first derive an extension of the Witten--Veneziano argument
(\cite{ff},\cite{ggg}) for the solution of the $U(1)$ problem, which
determines the
$O(1/N)$ correction to the relation $\sqrt{\chi^\prime(0)}/F_0
\vert_{N_{=\infty}} =
1$, where $F_0$ is the $\underline{\rm physical,~RG-invariant}$ $\eta^\prime$
singlet axial vector coupling to $j^{(0)}_{\mu_{5,inv}}$ (in the chiral limit).
Although the resulting correction tends indeed to suppress $\sqrt{\chi^\prime(0)}$
with respect to $F_0$, it still appears  to be a small perturbation on the $N =
\infty$ result; it is thus likely to be insufficient to account for the observed
suppression, at least as long as the nonet symmetry relation
$(F_0/F_8)\vert_{N_{=\infty}}=1$ remains approximately valid at $N = 3$ (the
normalization is such that $F_8 \simeq F_\pi / \sqrt{6}$). Therefore, assuming
$\sqrt{\chi^\prime(0)} / F_0\simeq 1$ the remaining possibility is that
there is a
large suppression of $F_0/F_8$ itself at finite $N$. I examine whether this
assumption is phenomenologically viable in Section 3, where I point out
that even a
moderate suppression of $F_0$ would lead to severe difficulties with the current
standard model \cite{hh} for $J/\psi\rightarrow (\eta,\eta^\prime )\gamma$
decays,
given the large $\eta - \eta^\prime$ mixing angle, which follows from an
analysis of
the octet electromagnetic (e.m.) sum rule for the $(\eta,\eta^\prime)\rightarrow
\gamma\gamma$ decays (too strong a suppression of $F_0$ is  not favoured
either by the
singlet e.m. sum rule). In Section 4,  I note that the observed smallness of
$G^{(0)}_{A,inv}$ might indicate a substantial glueball-like contribution to
$G^{(0)}_{A,inv}$, which should then cancel against that of the $\eta^\prime$,
assuming the latter to be of typical $G^{(0)}_{A\vert OZI}$ size if $F_0$ is not
suppressed (and could thus be identified to the quark spin piece
(\cite{jj}--\cite{lll}) of the nucleon in the chiral limit). I then draw a
parallel
with the occurrence of a sizeable violation of the $\eta - \eta^\prime$
saturation
hypothesis in the $\langle 0\vert \partial^\mu
j^{(0)}_{\mu_{5,inv}}\vert\gamma\gamma
\rangle
\vert_{q^2=0}$ matrix element.

\section{$\chi^\prime(0)$ at large $N$}

Consider the dispersion relation:
\beq
\chi (q^2) = \chi (0) + \chi^\prime(0) q^2 + {q^4\over\pi}  \int^\infty_{q^2_0}
{dq^{\prime 2}\over q^{\prime 2}}~~{Im \chi(q^{\prime 2})\over q^{\prime 2}
- q^2}
\label{6}
\eeq
where two subtractions are needed, since $\chi (q^2)$, which is of
dimension 4, is
$O(q^4)$ at large $q^2$. In the quarkless Yang-Mills theory, one can thus write
(symbolically):
\beq
\chi_{YM}(q^2) = A_{YM} + B_{YM} q^2 + {F^2_G M^4_G\over M^2_G - q^2} + \ldots
\label{7}
\eeq
where $F_G$ and $M_G$ are the coupling and mass of the lowest-lying
glueball state,
and the dots stand for more massive glueballs as well as continuum
contributions,
whereas in the presence of quarks, splitting out the $\eta^\prime$ contribution:
\beq
\chi (q^2) = {F^2_0 m^4_0\over m^2_0 - q^2} + \left[ A + B q^2 + {F^2_G
M^4_G \over
M^2_G - q^2} + \ldots \right]
\label{8}
\eeq
where $m_0$ is the $\eta^\prime$ mass in the chiral limit, and the subtraction
constants $(A_{YM}, B_{YM}), (A,B)$ (which are not reducible to the glueballs
contribution) have been introduced. Taking the $N \rightarrow\infty$ limit
in Eq.
(\ref{8}) at fixed $q^2\not= 0$, one then expects (since quark loops are
subleading
and decouple) $\chi (q^2) \rightarrow \chi_{YM}(q^2)$. Indeed, the $\eta^\prime$
contribution drops out, given that $F^2_0 = O(N)$, if one assumes \cite{ff}
$m^2_0 =
O(1/N)$, whereas the quantity within brackets in Eq. (\ref{8}) approaches
$\chi_{YM}(q^2)\vert_{N=\infty}$, i.e., $A\rightarrow A_{YM~\vert N=\infty}$,
$B\rightarrow B_{YM~\vert N=\infty}$ (and glueballs $\rightarrow$
glueballs$\vert_{N=\infty}$). The implication for $\chi^\prime(q^2)$ is
obtained by
expanding Eq. (\ref{8}) around $q^2 = 0$:
\bea
\chi (q^2) &=& \left[ F^2_0 m^2_0 + (A + F^2_G M^2_G + \ldots)\right] + q^2
\left[
F^2_0 + (B + F^2_G + \ldots )\right] + 0(q^4)\nonumber \\
&\equiv& \chi (0) + q^2 \chi^\prime (0) + O(q^4)~.
\label{9}
\eea
The basic QCD constraint (for massless quarks) $\chi (0) = 0$ then gives:
\beq
\chi (0) = F^2_0 m^2_0 + (A + F^2_G M^2_G + \ldots ) = 0~.
\label{10}
\eeq
Letting $N\rightarrow\infty$ in Eq. (\ref{10}), one first recovers the relation
(\cite{ff},\cite{ggg}):
\beq
F^2_0 m^2_0\vert_{N=\infty} = -(A_{YM} + F^2_G M^2_G + \ldots)\vert_{N=\infty}
\equiv -\chi_{YM}(0)\vert_{N=\infty}~,
\label{11}
\eeq
whereas for $\chi^\prime(0)$ one obtains from Eq. (\ref{9}) the additional
relation\footnote{This relation was first discovered in Ref. \cite{mm}, where it
was (interestingly) suggested by a QCD sum rule analysis of $\chi_{YM}(0)$.}:
\beq
(\chi^\prime(0) - F^2_0)\vert_{N=\infty} = (B_{YM} + F^2_G + \ldots
)\vert_{N=\infty} \equiv \chi^\prime_{YM}(0)\vert_{N=\infty}~.
\label{12}
\eeq
Since $\chi^\prime_{YM}(0)$ is $O(1)$ and $F^2_0$ is $O(N)$, Eq. (\ref{12})
requires $\chi^\prime(0)$ to be $O(N)$ and positive, in order that a
cancellation
takes place with $F^2_0$. On the other hand, a lattice calculation \cite{nn}, in
agreement with a QCD sum rule analysis \cite{mm}, yields
$\chi^\prime_{YM}(0) < 0$.
Writing Eq. (\ref{12}) as:
\beq
{\chi^\prime(0) \over F^2_0}~~~ {\simeq \atop N\rightarrow\infty}~~~ 1 +
{\chi^\prime_{YM}(0)\over F^2_0}
\label{13}
\eeq 
one thus finds the second term on the right-hand side gives the $O(1/N)$
correction
to the OZI limit relation $\chi^\prime (0)/F^2_0\vert_{N=\infty} = 1$, and
indeed
tends to suppress $\chi^\prime (0)$ with respect to $F^2_0$, since
$\chi^\prime_{YM}(0) < 0$. However, the correction appears numerically
small (from
Refs. \cite{mm} and \cite{nn} one gets $-\chi^\prime_{YM}(0)/F^2_0 \simeq 0.1$),
which suggests that the OZI violations in $\chi^\prime (0)/F^2_0$ are
probably small
and that the large-$N$ expansion is reliable for this ratio. In the next
section, I
investigate whether the assumption that there are instead large OZI violations
that strongly suppress the ratio $F_0/F_8$ at $N = 3$ is phenomenologically
viable.

The results of this section suggest a simple model for the structure
of $\chi(q^2)$ $\underline{\rm at~ finite~ N}$ \footnote{I am indebted to
G. Veneziano
for stressing this point.} in the presence of massless quarks, where it is
written as
the sum of the
$\eta^\prime$ pole contribution and the Yang--Mills topological susceptibility:
$\chi (q^2) = \chi_{YM}(q^2) + F^2_0m^4_0/(m^2_0-q^2)$, and $\chi_{YM}(q^2)$ is
further approximated by dropping the glueballs contribution, and keeping
only the
subtraction terms, namely taking:
$$
\chi_{YM}(q^2) \equiv A_{YM} + B_{YM}q^2 \equiv \chi_{YM}(0) +
\chi^\prime_{YM}(0)
q^2~.
$$
We thus get:
\beq
\chi (q^2) = \chi_{YM}(0) + \chi^\prime_{YM}(0) q^2 + {F^2_0 m^4_0\over
m^2_0 - q^2}~.
\label{14}
\eeq
The constraint $\chi (0) = 0$ yields $\chi_{YM}(0) + F^2_0 m^2_0 = 0$ [cf. Eq.
(\ref{11})]. Since $\chi^\prime (0) = F^2_0 + \chi^\prime_{YM} (0)$ [cf. Eq.
(\ref{12})], Eq. (\ref{14}) then becomes, after eliminating $m^2_0$:
\beq
\chi (q^2) = q^2 \left({\chi_{YM}(0)\over q^2 + {\chi_{YM}(0)\over F^2_0}} +
\chi^\prime_{YM}(0)\right)~.
\label{15}
\eeq
If $\chi^\prime_{YM}(0)$ is dropped, i.e. if one assumes $\chi_{YM}(q^2) \equiv
\chi_{YM}(0)$, one recovers an ansatz given in Ref. \cite{oo}, which yields
$\chi^\prime(0) = F^2_0$; the additional term $\chi^\prime_{YM}(0)$ accounts for
the OZI violation in this model.

\section{Implication of a small $F_0$ for $J/\psi\rightarrow (\eta
,\eta^\prime)\gamma$ and $(\eta , \eta^\prime )\rightarrow\gamma\gamma$ decays}

An analysis \cite{pp} of $\eta - \eta^\prime$ mixing using the anomalous Ward
identities does indicate that a large suppression of $F_0/F_8$ is indeed
possible
at large mixing angles, and at least appears to favour a moderate suppression in
this region (these results may, however, be changed by taking into account
\cite{qq} the recently calculated \cite{rr} $O(m^2_q)$ quark mass
corrections at $N
= \infty$). A large mixing angle is itself supported \cite{pp} by the data
on $(\eta
, \eta^\prime)\rightarrow\gamma\gamma$. However, even a modest suppression
of $F_0$
is in strong disagreement with the current standard model \cite{hh} for
$J/\psi\rightarrow (\eta , \eta^\prime )\gamma$. The argument (\cite{pp},
\cite{qq}) can be summarized as follows. From the octet e.m. anomaly sum rule
(assuming $\eta - \eta^\prime$ saturation):
\beq
F_{8\eta} A(\eta\rightarrow\gamma\gamma ) + F_{8\eta^\prime}
A(\eta^\prime\rightarrow\gamma\gamma ) = {1\over\sqrt{3}}
\label{16}
\eeq
[where $F_{8p} (p = \eta , \eta^\prime )$ are the couplings to $j^{(8)}_{\mu_5}$],
one can extract $F_{8\eta^\prime}$, using	 as input the experimentally
determined
amplitudes \cite{ss}: $A(\eta\rightarrow\gamma\gamma ) = (0.993 \pm
0.030)F^{-1}_\pi$ and $A(\eta^\prime\rightarrow\gamma\gamma ) = (1.280 \pm
0.085)
F^{-1}_\pi$, as well as the crucial perturbation theory estimate \cite{tt}:
$F_{8\eta}/F_\pi = 1.3 \pm 0.05$. As an indication, using $F_{8\eta}/F_\pi
= 1.25$,
one gets $\sin\theta\equiv -F_{8\eta}^\prime /F_\pi = 0.52$, a rather large
value.

Furthermore, the singlet couplings $F_{0p}$ can be constrained with the $J/\psi
\rightarrow (\eta , \eta^\prime )\gamma$ decays. Indeed the current
standard model
\cite{hh} for the ratio $\Gamma (J/\psi\rightarrow\eta^\prime\gamma )/\Gamma
(J/\psi\rightarrow\eta\gamma )$ relates it to $\tilde f_{\eta^\prime}/\tilde
f_\eta$ (where the $\tilde f_p$'s are the anomalous divergence couplings:
$\langle 0\vert Q\vert p\rangle = \tilde f_p m^2_p$):
\beq
R \equiv {\tilde f_{\eta^\prime}\over\tilde f_\eta} \simeq \left[{\Gamma
(J/\psi\rightarrow\eta^\prime\gamma )\over \Gamma
(J/\psi\rightarrow\eta\gamma)}~~{(M^2_{J/\psi} - m^2_\eta)^3\over
(M^2_{J/\psi}-m^2_{\eta^\prime})^3}\right]^{1/2}~~{m^2_\eta\over
m^2_{\eta^\prime}}~;
\label{17}
\eeq
This gives, using \cite{ss} $\Gamma (J/\psi\rightarrow\eta^\prime\gamma ) /
\Gamma (J/\psi\rightarrow\eta\gamma ) = 5.0 \pm 0.6$~:~~ $R_{\exp} = 0.81
\pm 0.05$.
But $\tilde f_p$ can be expressed \cite{pp} in terms of the corresponding axial
vector couplings $F_{ip}$ and the quark mass ratios $\beta / \gamma$ and $\gamma
/\alpha$:
\bea
{\tilde f_{\eta^\prime}\over F_\pi} &\equiv & {F_{0\eta^\prime}\over F_\pi} -
{f_{0\eta^\prime}\over F_\pi} \simeq {F_{0\eta^\prime}\over F_\pi} -
{\beta\over\gamma}~~{F_{8\eta^\prime}\over F_\pi} \nonumber \\ &&\nonumber \\
{\tilde f_\eta\over F_\pi} &\equiv & {F_{0\eta}\over F_\pi} -
{f_{0\eta}\over F_\pi}
\simeq {F_{0\eta}\over F_\pi} - {\gamma\over\alpha}~~{F_{8\eta}\over F_\pi}
\label{18}
\eea
where the $f_{0p}$'s are the ``na\"\i ve divergence" couplings,
$\alpha\equiv (2/3)
(m_u + m_d + 4m_s)$,
$\beta
\equiv (4/3) (m_u + m_d + m_s)$,
$\gamma\equiv -\sqrt{2} (\alpha -\beta)$, and we have the estimates 
\cite{tt} $\beta
/
\gamma \simeq 0.79$ and $\gamma /\alpha \simeq -0.67$. I shall simply use
the value
(obtained by putting $m_u = m_d = 0$): $\beta /\gamma = \gamma /\alpha =
-1/\sqrt{2}$. Then one gets:
\beq
{\tilde f_{\eta^\prime}\over \tilde f_\eta} \simeq {{F_{0\eta^\prime}\over
F_\pi} -
{\beta\over\gamma}~~{F_{8\eta^\prime}\over F_\pi} \over {F_{0\eta}\over F_\pi} -
{\gamma\over\alpha}~~{F_{8\eta}\over F_\pi} } = { {F_{0\eta^\prime}\over
F_\pi} -
{1\over\sqrt{2}}~~ \sin\theta \over {F_{0\eta}\over F_\pi} + {1\over \sqrt{2}}
~~{F_{8\eta}\over F_\pi} }~.
\label{19}
\eeq
Assuming again $F_{8\eta}/F_\pi$ = 1.25 and $\sin\theta \equiv
-F^\prime_{8\eta}/F_\pi$
= 0.52 from the octet sum rule, and taking $\tilde f_{\eta^\prime}/\tilde
f_\eta =
R_{\exp} \simeq 0.81$, Eq. (\ref{19}) then fixes $F_{0\eta}$ as a function of
$F_{0\eta^\prime}$, and one finds that unrealistically small values of
$F_{0\eta}$ are
required to fit $R_{\exp}$. For instance, assuming the moderately
suppressed value 
$F_{0\eta^\prime}/F_\pi$ = 0.90, Eq. (\ref{19}) gives 
$F_{0\eta}/F_\pi = -0.23$, which violates, even in sign, the large-$N$
expectation
\cite{pp}: $F_{0\eta} \simeq -F_{8\eta^\prime}$ ! Also, one still gets
$F_{0\eta} =
0$ even for $F_{0\eta^\prime}/F_\pi$ as large as 1.1. Clearly, the model of Eq.
(\ref{17}) is
$\underline{\rm incompatible~with~any~kind~of}$
$\underline{\rm suppression~whatsoever~of~the
~singlet~coupling}$ $F_{0\eta^\prime}$, given the large input values of the
octet couplings $F_{8\eta}$ and $-F_{8\eta^\prime}$. Since the quark mass
corrections that relate 
$F_{0\eta^\prime}$ to its chiral limit $F_0$ are small (they have been estimated
\cite{pp} to be $F_{0\eta^\prime} \simeq 1.16 F_0$), this observation probably
rules out the possibility that $F_0 / F_8 \simeq F_0 /F_\pi$ be substantially
suppressed.
In fact, for $F_{0\eta}/F_\pi = 0.50 (\simeq - F_{8\eta^\prime} /F_\pi)$, Eq.
(\ref{19}) gives $F_{0\eta^\prime} / F_\pi = 1.49$, hence $F_0 /F_\pi =
1.28$, an
$\underline{\rm enhancement}$! On the other hand, the singlet e.m. sum rule 
($\underline{\rm assuming~again}$ $\eta, \eta^\prime$ $\underline{\rm
saturation}$): 
\beq
F_{0\eta} A(\eta\rightarrow\gamma\gamma ) + F_{0\eta^\prime}
A(\eta^\prime\rightarrow\gamma\gamma ) = 2 \sqrt{2\over 3}
\label{20}
\eeq
does favour a ($\underline{\rm moderate})$\footnote{That is,  too strong a
suppression, such as the one needed to explain $G_{A,inv}^{(0)}$, is not
favoured
either by Eq.~(\ref{20}), which would then lead to values of $F_{0\eta}$
too large,
typically $F_{0\eta}/F_\pi \simeq F_{0\eta^\prime}/F_\pi \simeq 0.7$ !}
suppression
of $F_{0\eta^\prime}$, e.g., if one again assumes  $F_{0\eta}/F_\pi = 0.50$, one
deduces from Eq. (\ref{20})  $F_{0\eta^\prime}/F_\pi \simeq 0.89$ (this
suppression
would not be sufficient anyway to explain the magnitude of $G^{(0)}_{A,inv}$).
However, Eq. (\ref{19}) then gives (taking the same values as above for
$F_{8\eta}$
and $F_{8\eta^\prime}$) $R \simeq$ 0.38, still a factor of 2 below
$R_{\exp}$, in
accordance with the previous remarks (this potential conflict between the
singlet
e.m. sum rule and $R_{\exp}$ is further commented upon in the next section).

\section{On gluonic contributions to $\langle 0\vert\partial^\mu
j^{(0)}_{\mu_5}\vert
N\bar N\rangle$ and $\langle 0\vert\partial^\mu j^{(0)}_{\mu_5}\vert
\gamma\gamma\rangle$}

The smallness of $G^{(0)}_{A,inv}$ may alternatively be seen as the result of a
cancellation \cite{uu} between the $\eta^\prime$ and the (glueball + continuum)
contributions (I consider for simplicity the chiral limit, where the $\eta$
decouples from $G_A^{(0)}$). This picture can be given a precise content by
using
the invariant definition of the singlet current (which removes [see also
below] the
inconsistencies with renormalization group invariance discussed in Refs.
\cite{ee} and
\cite{vv}). One can define, splitting out the $\eta^\prime$ contribution
($g_{\eta^\prime NN}$ is the $\eta^\prime$-nucleon coupling):
\beq
\langle 0\vert\partial^\mu j^{(0)}_{\mu_5,inv}\vert N\bar N\rangle \propto
\Delta\Sigma_{inv} \equiv G_{A,inv}^{(0)} \equiv F_0~ g_{\eta^\prime NN} + 
\Delta\Gamma_{inv}~,
\label{21}
\eeq
where $\Delta\Gamma_{inv}$ represents the (glueball + continuum)
contribution, and
all quantities in Eq. (\ref{21}) are renormalization group-invariant. It is then
attractive to identify the ``quark contribution"
$\Delta\Sigma^\prime_{inv}$ to the
nucleon spin with the $\eta^\prime$ contribution\footnote{In the presence
of $SU(3)$
breaking, one should also add the $\eta$ contribution.} $F_0~
g_{\eta^\prime NN}$,
while $\Delta\Gamma_{inv}$ would represent the ``gluon contribution"
({\cite{jj}--\cite{lll}). If $F_0$ is indeed not suppressed, one might further
assume, in the line of the latter references, that
\beq
\Delta\Sigma^\prime_{inv} = F_0 ~g_{\eta^\prime NN} \sim (F_0~ g_{\eta^\prime
NN})\vert_{OZI} = G_A^{(0)}\vert_{OZI}
\label{22}
\eeq
and attribute the small value of $ G_{A,inv}^{(0)}$ to the effect of a
substantial
(negative) $\Delta\Gamma_{inv}$, i.e. $G_{A,inv}^{(0)} \simeq
G_{A\vert_{OZI}}^{(0)} + \Delta\Gamma_{inv}$ (it could also be that both $F_0~
g_{\eta^\prime NN}$ and $\Delta\Gamma_{inv}$ are suppressed, in which case $F_0~
g_{\eta^\prime NN}$ would still differ from $\Delta\Sigma^\prime_{inv}$ by some
additional, non-perturbative (\cite{ww}, \cite{bb}) contributions, e.g. $F_0~
g_{\eta^\prime NN} = \Delta\Sigma^\prime_{inv} - N_f \Omega$)
\footnote{Alternatively
one could have $\Delta\Sigma^\prime_{inv} = F_0~ g_{\eta^\prime NN}$ small, with
$g_{\eta^\prime NN}$ suppressed, as suggested by the Skyrme model
\cite{yy}.}.  Such a
proposal, which identifies
$\Delta\Gamma_{inv}$ to
$G_{A,inv}^{(0)} -  F_0~ g_{\eta^\prime NN}$, is complementary to the
QCD-improved
parton model approach of Refs.~(\cite{jj}-\cite{lll}}), which starts from
the implicit
assumption that it is possible to define a (perturbatively wise \cite{zz})
unique
physical gluon distribution, which could independently be measured in
suitable hard
processes.

There is an interesting parallel with the situation for the
$\langle 0\vert \partial^\mu j^{(0)}_{\mu_5,inv}
\vert\gamma\gamma\rangle\vert_{q^2=0}$
matrix element: the above-mentioned conflict between the singlet e.m. sum
rule (which
favours a moderately suppressed $F_0$) and $R_{\exp}$ (which favours an
unsuppressed, or even enhanced, $F_0$) may be resolved by assuming that $F_0$ is
indeed not suppressed. The resulting discrepancy in the singlet sum rule
(where the
$\eta$ and $\eta^\prime$ contribution now by itself exceeds the right-hand side)
can then be attributed to a substantial (negative) gluon-like contribution,
as in
the nucleon channel. Actually, considering the standard $\mu$-dependent
renormalization of the singlet current, Eq. (\ref{2}), one may suspect that an
independent subtraction constant $\Delta_0(\mu )$ enters the sum rule,
$\underline{\rm in~addition~to~the~glueball~contribution}$, i.e. Eq. (\ref{20})
should be replaced by:
\bea
\langle 0\vert\partial^\mu j^{(0)}_{\mu_5}(\mu ) \vert\gamma\gamma
\rangle\vert_{q^2=0}
&&\propto F_{0\eta}(\mu ) A(\eta\rightarrow\gamma\gamma )\nonumber \\
\nonumber \\
&& + F_{0\eta^\prime}(\mu )
A(\eta^\prime\rightarrow\gamma\gamma ) + \left[ F_G(\mu )
A(G\rightarrow\gamma\gamma )
+ \ldots \right] \nonumber \\
&&+ \Delta_0(\mu ) \equiv 2\sqrt{2\over 3}~.
\label{23}
\eea
The introduction of $\Delta_0(\mu )$ appears necessary\footnote{The origin of
$\Delta_0(\mu )$ can probably be traced back to the additional ultraviolet
divergence in $\langle 0\vert\partial^\mu
j^{(0)}_{\mu_5}\vert\gamma\gamma\rangle$
arising first at the $O(\alpha^2_s\alpha )$ level from the (lowest-order in
electromagnetism)
$O(\alpha )$ coupling to $\gamma\gamma$,  which is distinct from the standard
one responsible for the QCD anomalous dimension of $j^{(0)}_{\mu_5}$ due to
the strong
anomaly.} to resolve the conflict \cite{vv} between the multiplicative
renormalizability of all the singlet axial vector couplings in the
left-hand side
of Eq. (\ref{23}), and the $\mu$-independence of the right-hand side (then,
letting
$\mu\rightarrow\infty$ one recovers the equation written in terms of the
invariant
couplings, with $\Delta_0(\mu )\rightarrow \Delta_{0,inv}$). This
$\Delta_0$ is also
welcome to explain the conjectured existence of a sizeable discrepancy in the
singlet sum rule with respect to the $\eta , \eta^\prime$ saturation hypothesis,
since the glueballs by themselves are expected to couple too weakly to the
photons
to be responsible for the entire discrepancy ($\Delta_0$ also recalls the
necessary subtraction constant   (\cite{ff},\cite{pp}) needed to cancel 
[see Eq.
(\ref{10})] the $\eta^\prime$ contribution and implement the constraint $\chi
(q^2=0)=0$ in the chiral limit of QCD; a subtraction constant has, however,
no reason
to be present in $\langle 0\vert\partial^\mu j^{(0)}_{\mu_5}\vert N\bar
N\rangle$).

To conclude, the present analysis offers only scarce evidence for the
suppression
of $\chi^\prime(0)$ as implied by Eq. (\ref{3}). Although $\chi^\prime
(0)/F^2_0$
is indeed suppressed at next-to-leading order in $1/N$, the correction appears
small, and of typical perturbative (in $1/N$) size. Furthermore, $F_0$
itself does
not seem to be suppressed on phenomenological grounds, compared with $F_8$
(and may even
turn out to be $\underline{\rm predicted}$ enhanced at large $\sin\theta$
once the
$O(m^2_q$) corrections are taken into account in the Ward identity analysis of
$\eta - \eta^\prime$ mixing (\cite{pp},\cite{qq}). Assuming that $F_0$ is not
suppressed, an intriguing picture of large (negative) gluon-like
contributions to
$\langle 0\vert\partial^\mu j^{(0)}_{\mu_5}\vert N\bar N\rangle\vert_{q^2=0}$, 
$\langle 0\vert\partial^\mu
j^{(0)}_{\mu_5}\vert\gamma\gamma\rangle\vert_{q^2=0}$ and
$\chi (q^2=0)$ tentatively emerges.

\vspace*{1cm}
\noindent
{\bf Acknowledgements}

	I thank the CERN Theoretical Physics Division for the hospitality, and G.
Altarelli, P.~Ball, J.-M. Fr\`ere, H. Leutwyler and G. Veneziano for clarifying
discussions and exchange of information.

\vfill\eject

\end{document}